\documentclass[10pt]{article}
\usepackage[]{amsmath,amsfonts,amsthm,amssymb}
\title{Classicality and connectedness for state property systems and
closure spaces\footnote{Accepted for publication in {\it International Journal of Theoretical Physics}}}
\author{Diederik Aerts, Didier Deses and Ann Van der Voorde}
\date{}
\newtheorem{theorem}{Theorem}
\newtheorem{definition}{Definition}

\newcommand{\cat}[1]{{\bf #1}}
\begin{document}
\maketitle
\centerline{FUND and TOPO,}
\centerline{Department of Mathematics, Brussels Free University,}
\centerline{Pleinlaan 2, B-1050 Brussels, Belgium}
\centerline{{\tt diraerts,diddesen,avdvoord@vub.ac.be}}

\begin{abstract}
\noindent
In \cite{A4,ACVdVVS} a physical entity is described by means of a state property
system and it is shown that there is an categorical equivalence
between the category $\cat{SPS}$ of state property systems and the category
$\cat{Cl}$ of closure spaces. In this note we prove, using this
equivalence between categories, that the concept of connectedness for closure
spaces can be used to formulate a decomposition theorem, which allows us to
split a state property system into a number of `pure nonclassical
state property systems' and a `totally classical state property
system'.
\end{abstract}

\noindent {\bf AMS Classification:} 81B10 \\ {\bf Keywords:}
property lattice, classical property, state property system, pure
nonclassical state property system, closure space, connectedness

\section{State property systems and closure spaces}

In \cite{A4} a physical entity $S$ is described by means of a set of states
$\Sigma$, a complete lattice of properties $\mathcal{L}$ and a map
$\xi:\Sigma\to \mathcal{P}(\mathcal{L})$. For any state $p\in \Sigma$ of the
entity $\xi(p)$ is the set of all properties which are actual whenever $S$ is
in the state $p$. Since $\mathcal{L}$ is a complete lattice it is partially
ordered, with the physical meaning of the partial order relation $<$ being the
following: $a, b \in {\cal L}$, such that $a < b$ means that whenever property
$a$ is actual for the entity $S$, also property $b$ is actual for the entity
$S$. If $\mathcal{L}$ is a complete lattice, it means that for an
arbitrary family of properties $(a_i)_i \in {\cal L}$ also the infimum
$\wedge_ia_i$ of this family is a property. The property $\wedge_ia_i$ is
the property that is actual if and only if all of the properties $a_i$ are
actual. Hence the infimum represents the logical `and'. The minimal element $0$
of the lattice of properties is the property that is never actual ({\it e.g.}
the physical entity does not exist). Using the intuitive idea of what a physist
means by a subsystem, a notion of morphism between state property systems was
introduced in \cite{ACVdVVS}, in order to form the category of state property
systems.

\begin{definition}
A triple $(\Sigma,\mathcal{L},\xi)$ is called a state property system if
$\Sigma$ is a set, $\mathcal{L}$ is a complete lattice (with top $I$ and
bottom $0$) and $\xi : \Sigma \rightarrow \mathcal{P}(\mathcal{ L})$ is a
function such that for $p \in \Sigma$ and $(a_i)_i, a, b \in
\mathcal{L}$, we have: \begin{itemize}
\item $0 \not\in \xi(p)$
\item $a_i \in \xi(p)\ \forall i \Rightarrow \wedge_i a_i \in \xi(p)$
\item $a < b \Leftrightarrow \forall r \in \Sigma:a \in \xi(r) \text{ then } b
\in \xi(r)$
\end{itemize}
Suppose that $(\Sigma,\mathcal{L},\xi)$ and $(\Sigma',\mathcal{L}',\xi')$ are
state property systems then $(m,n):(\Sigma',\mathcal{L}',\xi') \to
(\Sigma,\mathcal{L},\xi)$ is called an $\cat{SPS}$-morphism if $m :
\Sigma' \to \Sigma$ and $n : \mathcal{L} \to \mathcal{L}'$ are functions
such that for $a \in \mathcal{L}$ and $p' \in \Sigma'$:
$$ a \in \xi(m(p')) \Leftrightarrow n(a) \in \xi'(p')$$
The category of state property systems and their morphisms is
denoted by $\cat{SPS}$.
\end{definition}

If $(\Sigma,\mathcal{L},\xi)$ is a state property system then its Cartan map is
the mapping  \mbox{$\kappa:\mathcal{L} \to \mathcal{P}(\Sigma)$} defined by :
$$\kappa: \mathcal{L} \rightarrow \mathcal{P}(\Sigma):a \mapsto \kappa(a) =
\{p\in \Sigma \ \vert\ a \in \xi(p)\}$$
We also introduce the map $s_{\xi}$ which maps a state $p$ to the strongest
property it makes actual, i.e.
$$s_{\xi} : \Sigma \to \mathcal{L} : p \mapsto \wedge \xi(p)$$

It was amazing to be able to prove (see \cite{ACVdVVS}) that the category of
states property systems and its morphisms is equivalent to the category of
closure spaces and the continuous
maps.

\begin{definition}
A closure space $(X,\mathcal{F})$ is a set $X$ and a collection of `closed'
subsets $\mathcal{F}$ such that $\emptyset \in\mathcal{F}$ and $(F_i)_i \in
\mathcal{F}$ implies $\cap_iF_i \in \mathcal{F}$.  If $(X,\mathcal{F})$ and
$(Y,\mathcal{G})$ are closure spaces then a function $f : (X,\mathcal{F}) \to
(Y,\mathcal{G})$ is called continuous if $\forall B \in \mathcal{G}:f^{-1}(B)
\in \mathcal{F}$. The category of closure spaces and continuous functions is
denoted by $\cat{Cl}$. \end{definition}

\begin{theorem}\cite{ACVdVVS}\label{thm:equiv}
The following functors define an equivalence of categories.
\begin{eqnarray*}
F:\cat{SPS} &\rightarrow& \cat{Cl} \\
(\Sigma,\mathcal{L},\xi)&\mapsto& F(\Sigma,\mathcal{L},\xi) = (\Sigma,\kappa(\mathcal{L}))\\
(m,n) &\mapsto& m
\end{eqnarray*}
and
\begin{eqnarray*}
G:\cat{Cl}&\rightarrow& \cat{SPS} \\
(X,\mathcal{F})&\mapsto&G(X,\mathcal{F})=(X,\mathcal{F},\bar\xi)\\
m &\mapsto&(m,m^{-1})
\end{eqnarray*}
where $\bar\xi : X \rightarrow \mathcal{P}(\mathcal{F}) :  p \mapsto
\{F \in \mathcal{F}\ \vert\ p \in F\}$.
\end{theorem}

\section{Super selection rules}

In this section we start to distuinguish the classical aspects of the
structure from the quantum aspects. The superposition states are the
states that do not exist in classical physics and hence their
appearance is one of the important quantum aspects. The concept of
superposition can be traced back within this general setting, by introducing
the idea of `superselection rule'. Two properties are separated by a
superselection rule if and only if there do not exist `superposition states'
related to these two properties.

\begin{definition}
Consider a state property system $(\Sigma,\mathcal{L},\xi)$. For $a,
b \in \mathcal{L}$ we say that $a$ and $b$ are separated by a super
selection rule, and denote $a$ ssr $b$, if and only if for $p \in \Sigma$ we
have:
$$ a \vee b \in \xi(p) \Rightarrow a \in \xi(p) {\rm \ or}\ b \in\xi(p)$$
\end{definition}

It is easy to show that $a\ ssr\ b$ if and only if $\kappa(a\vee
b)=\kappa(a)\cup\kappa(b)$. Hence the following theorem.

\begin{theorem}
Consider a state property system $(\Sigma,\mathcal{L},\xi)$ and its
corresponding closure space $(\Sigma,\kappa(\mathcal{L}))$. Then
$(\Sigma,\mathcal{L},\xi)$ is classical, in the sense that every two properties
of $\mathcal{L}$ are separated by a super selection rule, if and only if
$(\Sigma,\kappa(\mathcal{L}))$  is a topological space.
\end{theorem}

We are ready now to introduce the concept of a `classical
property'.

\begin{definition}\label{def:class}
Consider a state property system $(\Sigma,\mathcal{L},\xi)$. We say
that a property $a \in \mathcal{L}$ is a `classical property', if
there exists a property $a^c \in \mathcal{L}$ such that $a \vee a^c =
I$, $a \wedge a^c = 0$ and $a$
ssr $a^c$. \end{definition}

We recall that a closure space $(X,\mathcal{F})$ is called connected if the
only clopen (i.e. closed and open) sets are $\emptyset$ and $X$. We shall see
now that these subsets, which make closure systems disconnected, are exactly
the subsets corresponding to classical properties.

\begin{theorem}
\label{thm:classclopen}
Consider a state property system $(\Sigma,\mathcal{L},\xi)$ and its
corresponding closure space $(\Sigma,\kappa(\mathcal{L}))$. For $a \in
\mathcal{L}$ we have: $$a \text{ is classical }\Leftrightarrow \kappa(a) \text{
is clopen}$$ \end{theorem}

\begin{proof}
If $a$ is classical $a\ ssr\ a^c$, so we have $\Sigma=\kappa(a\vee
a^c)=\kappa(a)\cup \kappa(a^c)$ and $\emptyset=\kappa(a\wedge
a^c)=\kappa(a)\cap \kappa(a^c)$. Therefore $\kappa(a)$ is clopen.
Conversely if $\kappa(a)$ is clopen we choose $a^c$ such that
$\kappa(a^c)=\Sigma\setminus \kappa(a)$. Obviously $a$ and $a^c$ satisfy the
conditions of Definition \ref{def:class}. \end{proof}

\begin{definition}
A state property system $(\Sigma,\mathcal{L},\xi)$ is called a pure nonclassical state
property system if the properties $0$ and $I$ are the only
classical properties.
\end{definition}

With this definition and Theorem \ref{thm:classclopen} it is easy to prove the
following.

\begin{theorem}
Consider a state property system $(\Sigma,\mathcal{L},\xi)$ and its
corresponding closure space $(\Sigma,\kappa(\mathcal{L}))$. Then
$(\Sigma,\mathcal{L},\xi)$ is a pure nonclassical state
property system if and only if $(\Sigma,\kappa(\mathcal{L}))$ is a connected
closure space.
\end{theorem}

\section{Decomposition theorem and the classical part of a state property
system}

As for topological spaces, every closure space can be decomposed
uniquely into connected components. In the following we say that,
for a closure space $(X,\mathcal{F})$, a subset $A \subseteq X$ is
connected if the induced subspace is connected. It can be shown
that the union of any family of connected subsets having at least
one point in common is also connected. So the component of an
element $x \in X$ defined by
$$K_{\cat{Cl}}(x)=\bigcup \{A\subseteq X \ | \ x \in A, A \text{ connected }\}$$
is connected and therefore called the connection component of $x$. Moreover, it
is a maximal connected set in $X$ in the sense that there is no connected
subset of $X$ which properly contains $K_{\cat{Cl}}(x)$. From this it follows
that for closure spaces $(X,\mathcal{F})$ the set of all distinct connection
components in $X$ form a partition of $X$. So we can consider the following
equivalence relation on $X$ : for $x,y \in X$ we say that $x$ is equivalent
with $y$ if and only if the connection components $K_{\cat{Cl}}(x)$ and $K_{\cat{Cl}}(y)$
are equal. Further we remark that the connection components are closed sets.
A closure space is called totally disconnected if for each $x\in X$,
$K_{\cat{Cl}}(x)=\{x\}$.

In the following we will try to decompose state property systems similarly into
different components. Suppose $(\Sigma,\mathcal{L},\xi)$ is a state property
system, then the corresponding closure space $(\Sigma,\kappa(\mathcal{L}))$ has
a unique partition $\Omega$ into maximal connected subsets. Each of these
subsets induces of course a connected closure space, hence one can use
the equivalence from Theorem \ref{thm:equiv} in order to obtain a number of
pure non-classical state property systems $G(\omega,\mathcal{F}_{\omega})$,
where $\mathcal{F}_{\omega}=\{\kappa(a)\cap \omega|a\in\mathcal{L}\}$ and
$\omega\in \Omega$. It is easily seen that this construction can be used
directly in terms of the given state property system as follows.

\begin{theorem}
Let $(\Sigma,\mathcal{L},\xi)$ be a state property system and let $(\Sigma,\kappa(\mathcal{L}))$
the corresponding closure space. Consider the following equivalence relation on
$\Sigma$ :
$$p \sim q \Leftrightarrow K_{\bf Cls}(p)=K_{\bf Cls}(q)$$
with equivalence classes $\Omega=\{\omega(p)|p\in \Sigma\}$.
If $\omega \in \Omega$ we define the following:
\begin{eqnarray*}
\Sigma_\omega &=&\omega  = \{p \in \Sigma \ | \ \omega(p) = \omega\} \\
s(\omega)&=&s(\omega(p))=a, \text{ such
that } \kappa(a)=\omega(p) \\ \mathcal{L}_\omega&=&[0,s(\omega)]=\{a\in
\mathcal{L} \ | \ 0 \leq a \leq s(\omega) \}\subset \mathcal{L}\\ \xi_\omega
&:&\Sigma_\omega\to \mathcal{P}({\cal L}_\omega):p \mapsto \xi(p)\cap
\mathcal{L}_\omega \end{eqnarray*}
Then $(\Sigma_\omega,\mathcal{L}_\omega,\xi_\omega)$ is a pure
non-classical state property system for each $\omega\in\Omega$.
\end{theorem}

Continuing on the same line of thought, one can make the following
construction, using classical methods from topology. Suppose
$(\Sigma,\mathcal{L},\xi)$ is a state property system as above, with
corresponding closure $(\Sigma,\kappa(\mathcal{L}))$ which has a unique
partition $\Omega$ into maximal connected subspaces. We take the quotient in
$\cat{Cl}$ induced by the canonical surjection $\omega:\Sigma\to \Omega$, i.e.
the coarsest closure structure $\mathcal{C}$ on $\Omega$ such that $\omega$ is
a continuous function. This yields a totally disconnected closure space
$(\Omega,\mathcal{C})$, which by means of the equivalence from Theorem
\ref{thm:equiv} corresponds to a state property system
$G(\Omega,\mathcal{C})$, which can be considered as `totally classical'.
An explicit construction is given as follows.

\begin{theorem}
Let $(\Sigma,\mathcal{L},\xi)$ be a state property system. If we introduce the
following :
\begin{eqnarray*}
\Omega&=&\{\omega(p)|p\in \Sigma\}\\
\mathcal{C}&=&\{\vee s(\omega_i)|\omega_i\in \Omega\} \\
\eta&:&\Omega \to {\cal P}(\mathcal{C}):\omega=\omega(p)\mapsto \xi(p)\cap
\mathcal{C} \end{eqnarray*}
then $(\Omega,{\cal C},\eta)$ is a totally classical state property
system, in the sense that the only pure nonclassical segments
(i.e. segments with no proper classical elements) are trivial,
i.e. $\{0,s(\omega)\}$.
\end{theorem}

Summarizing the previous results we get that any state property system
$(\Sigma,\mathcal{L},\xi)$ can be decomposed into:
\begin{itemize}
\item a number of pure nonclassical state property systems
$(\Sigma_\omega,{\cal L}_\omega,\xi_\omega),\omega \in \Omega$
\item and a totally classical state property system $(\Omega,\mathcal{C},\eta)$
\end{itemize}

In this last paragraph we want to show how it is possible to extract
the classical part of a state property system. Consider again a state property
system $(\Sigma,\mathcal{L},\xi)$ and its associated closure space
$(\Sigma,\kappa(\mathcal{L}))$. A closure space is called zero-dimensional if
every closed set can be written as an intersection of clopen subsets. Let
$\mathcal{C}'=\{\bigcap_i A_i| A_i \text{ clopen in }
(\Sigma,\kappa(\mathcal{L}))\}$, then $(\Sigma,\mathcal{C}')$ is a
zero-dimensional closure space. This closure space correspond to a state
property system $G(\Sigma,\kappa(\mathcal{L}))$, which can be considered as the
classical part of the $(\Sigma,\mathcal{L},\xi)$. It can also be described as
follows.

\begin{theorem}
For a state property system $(\Sigma,\mathcal{L},\xi)$, then we define
\begin{eqnarray*}
\mathcal{C}'&=&\{\wedge_i a_i| a_i \text{ is a classical property}\} \\
\xi'&:&\Sigma \to \mathcal{P}(\mathcal{C}'):p\mapsto \xi(p)\cap \mathcal{C}'
\end{eqnarray*}
$(\Sigma,\mathcal{C}',\xi')$ is a state property system which we shall refer to
as the classical part of $(\Sigma,\mathcal{L},\xi)$.
\end{theorem}

For a more complete overview concerning different notions of classicality in
the setting of state property system we refer to the book \cite{AD}.


\end{document}